\documentclass[aps,prl,twocolumn,superscriptaddress,floats,epsfig,showpacs]{revtex4-1}
\usepackage{bbm}
\usepackage{amssymb}
\usepackage{ifpdf}
\usepackage{color}
\usepackage{graphicx}
\usepackage{amsmath}
\usepackage{amsfonts}
\usepackage{hyperref}
\usepackage{dcolumn}
\usepackage{bm}
\usepackage{xcolor}
\usepackage{subfigure}
\usepackage{verbatim}

\begin{document}

\title{Quench Dynamics of Three-Dimensional Disordered Bose Gases:
Condensation, Superfluidity and Fingerprint of Dynamical Bose Glass}
\author{Lei Chen}
\affiliation{Shenyang National Laboratory for Materials Science, Institute
of Metal Research, Chinese Academy of Sciences, Wenhua Road, 72, Shenyang,
China}
\author{Zhaoxin Liang}
\email{Corresponding author: E-mail: zhxliang@imr.ac.cn}
\affiliation{Shenyang National Laboratory for Materials Science, Institute
of Metal Research, Chinese Academy of Sciences, Wenhua Road, 72, Shenyang,
China}

\author{Ying Hu}
\affiliation{Institute for Quantum Optics and Quantum Information of the
Austrian Academy of Sciences, A-6020 Innsbruck, Austria}
\author{Zhidong Zhang}
\affiliation{Shenyang National Laboratory for Materials Science, Institute
of Metal Research, Chinese Academy of Sciences, Wenhua Road, 72, Shenyang,
China}

\begin{abstract}
In an equilibrium three-dimensional (3D) disordered condensate, it's well established that disorder can generate an amount of normal fluid equaling to $\frac{4}{3}$ of the condensate depletion. The concept that the superfluid is more volatile to the existence of disorder than the condensate is crucial
to the understanding of Bose glass phase. In this Letter, we show that, by bringing a
weakly disordered 3D condensate to nonequilibrium regime via
a quantum quench in the interaction, disorder can destroy superfluid significantly
more, leading to a steady state in which the normal fluid density far
exceeds $\frac{4}{3}$ of the condensate depletion. This suggests a
possibility of engineering Bose Glass in the dynamic regime. As both the
condensate density and superfluid density are measurable
quantities, our results allow an experimental demonstration of the dramatized
interplay between the disorder and interaction in the nonequilibrium scenario.
\end{abstract}

\maketitle

Interaction and disorder are two basic elements in nature. Their competition
underlies many intriguing phenomena in the equilibrium physics, such as
Anderson localization \cite{Localization} and the emergence of Bose glass
phase \cite{DisorderBose}. Very recently, the remarkable experimental
progress of quenching ultracold Bose gas \cite{QE1} in tunable disordered
potentials \cite{Disorder1,Disorder2,Disorder3,Disorder3} has generated a surge of new 
interests in studying this old problem in the non-equilibrium regime, in
which the combined effects of disorder and interactions can become much more
dramatic. In this context, while a focus of theoretical research \cite
{Tavora2014,Errico2014} has been on how a disordered system relaxes after
a variety of quantum quench, we address below the problem of how to feasibly
illustrate the quench effects on the interplay between the disorder and
interaction in experimentally observable quantities.

The point of this work is to revisit two fundamental and measurable
quantities, condensate density \cite{ConceptsPines,
ConceptsBaym,ConceptPit,Xu} and the superfluid density \cite{Cooper2010,
Ho2009,Carusotto2011, Keeling}, in the new context of quench dynamics of a
disordered Bose-Einstein condensate (BEC) at three dimension (3D). In the
equilibrium regime, the different fate of this two quantities in the
presence of disorder has been crucial for the understanding of the Bose glass
state \cite{DisorderBose}. In the pioneering work \cite{Huang1992} on
disordered 3D BECs at their ground state, it is pointed out that the weak disorder can generate a
normal fluid $\rho _{n}$ equaling $\frac{4}{3}$ of the condensate depletion $%
\rho _{ex}$ even at zero temperature. This ratio has been later shown to be
quite generic, arising also in 3D BECs with both strong interaction and
strong disorder \cite{Lotatin2002}, as well as in external potentials \cite%
{DisorderBEC}. This has led to the speculation on the existence of Bose
glass, in which superfluid vanishes but finite condensate survives. The
present work shows that, by combining with the effect of quantum quench in
the interaction, disorder can destroy even more superfluid than the
condensate compared to the equilibrium case. In particular, a ratio $\rho
_{n}/\rho _{ex}>>4/3$ can be achieved in the steady state of a weakly
disordered 3D BEC after the interaction quench. The observation that the superfluid becomes increasingly volatile in the nonequiibrium scenario to the effect of disorder implies a possibility to
engineering the Bose glass in the dynamic regime, which we shall referred to
as the dynamical Bose glass. 

\textit{Model Hamiltonian.---} We consider a weakly interacting 3D Bose gas
in the presence of disordered potentials under a quantum quench in the
interaction. The corresponding second-quantized Hamiltonian reads \cite%
{Huang1992,Yukalov,YingHu2009,Gaul2011}
\begin{eqnarray}
H-\mu N&=&\int d^3\mathbf{r}\hat{\Psi}^{\dagger}(\mathbf{r})\Big[-\frac{%
\hbar^2\nabla^2}{2m}-\mu+V_{dis}\left(\mathbf{r}\right)  \notag \\
&+&\frac{1}{2}g(t)\hat{\Psi}^{\dagger}(\mathbf{r})\hat{\Psi}(\mathbf{r}) %
\Big]\hat{\Psi}(\mathbf{r}),  \label{Hamiltonian}
\end{eqnarray}
where $\hat{\Psi}(\mathbf{r})$ is the field operator for bosons with mass $m$%
, $\mu$ is the chemical potential, $N=\int d\mathbf{r \hat{\Psi}^{\dagger}(r)%
\hat{\Psi}}(\mathbf{r})$ is the number operator and $V_{dis}\left(\mathbf{r}%
\right)$ represents the disordered potential. The $g(t)$ in Hamiltonian (\ref%
{Hamiltonian}) describes the quench protocol for the interaction parameter.
Specifically, we consider the case when the system is initially prepared at
the ground state $|\Psi(0)\rangle$ of Hamiltonian (\ref{Hamiltonian}) with $%
g=g_i$ labeled by $H_i$; then, at $t=0$, the interaction strength is
suddenly switched to $g=g_f$ such that the time evolution from $t>0$ is
governed by the finial Hamiltonian (\ref{Hamiltonian}) of $H_f$.
Accordingly, we write
\begin{equation}
g(t)=g_i\left[1+\Theta(t)\left(\tilde{g}-1\right)\right],  \label{Quench}
\end{equation}
with $\tilde{g}=g_f/g_i$ and $\Theta (t)$ being the Heaviside function.
Experimentally, the interaction quench in Eq. (\ref{Quench}) can be achieved
by using Feshbach resonance \cite{Chin2010}.

For $V_{dis}\left(\mathbf{r}\right)$ in Hamiltonian (\ref{Hamiltonian}), we
consider its realization \cite%
{Disorder1,Disorder2,Disorder3,Huang1992,Lotatin2002,DisorderBEC,Yukalov,YingHu2009,Gaul2011}
via the random distribution of quenched impurity atoms described by $V_{dis}(%
\mathbf{r})=g_{imp}\sum_{i=1}^{N_{imp}}\delta \left(\mathbf{r}-\mathbf{r}%
_i\right)$ with $g_{imp}$ being the coupling constant of an impurity-boson
pair \cite{Huang1992}, $\mathbf{r}_i$ the randomly distributed positions of
the impurities, and $N_{imp}$ counting the number of $\mathbf{r}_i$. The
randomness is uniformly distributed and Gaussian correlated \cite{YingHu2009}%
, such that $\langle V_0\rangle=g_{imp}N_{imp}/V$ ($V$ is the system's volume)
and
\begin{equation}
R_0=\frac{1}{V}{\langle V_{-\mathbf{k}}V_{\mathbf{k}}\rangle},
\end{equation}
with $V_{\mathbf{k}}=(1/V)\int d\mathbf{r}e^{i\mathbf{k}\cdot\mathbf{r}%
}V_{dis}(\mathbf{r})$ and $\langle ... \rangle$ describing the ensemble
average over all possible realizations of disorder configurations.
\begin{figure}[t]
\centering
\includegraphics[width=9.0cm]{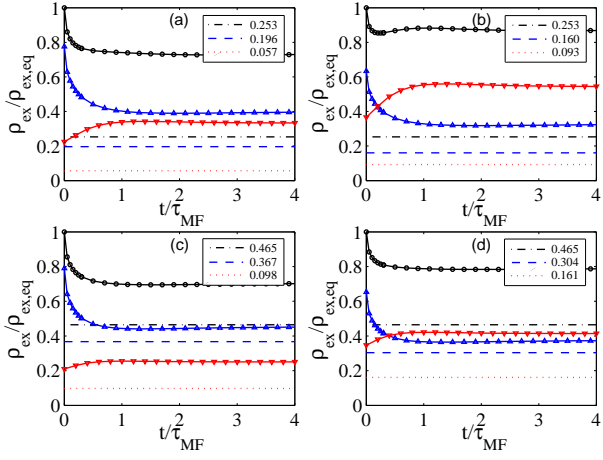}
\caption{(Color online) Dynamics of the quantum depletion $\protect\rho%
_{ex}(t)$ after the quench. The solid curves are the quantum depletion $%
\protect\rho_{ex}(t)$ redistribution of a disordered Bose gas following a
sudden change of interactions from $g_i$ to $g_f$ via the time. The dashed
curves are the excitation fraction at the final interaction strength in
equilibrium. The blue and red curves correspond to the interaction- and
disordered-induced quantum depletion, whereas the black curve describes the
quantum depletion due to the combined effects. The characteristic density
and relaxation time are set by $\protect\rho_{ex,eq}=1/3\protect\pi^2\protect%
\zeta^3+(1/4\protect\pi \protect\zeta^3)\tilde{R}_0$ and $\protect\tau%
_{MF}=\hbar/g_{f}\protect\rho_0$ respectively. Parameters are given for (a) $%
\tilde{R}_0=0.1$ and $\tilde{g}=0.4$; (b) $\tilde{R}_0=0.2$ and $\tilde{g}%
=0.4$; (c) $\tilde{R}_0=0.1$ and $\tilde{g}=0.6$; (d) $\tilde{R}_0=0.2$ and $%
\tilde{g}=0.6$.}
\label{Condensatedensity}
\end{figure}

We shall focus on the regime of both weak interaction and weak disorder, in
which Hamiltonian (\ref{Hamiltonian}) can be well described using the
standard Bogoliubov approximation \cite{Natu2013,Huang1992,YingHu2009} and the
resulting expression reads
\begin{eqnarray}
H_{\text{eff}}(t) &=&\sum_{\mathbf{k}\neq 0}\left( \epsilon _{\mathbf{k}%
}-\mu \right) \hat{a}_{\mathbf{k}}^{\dagger }\hat{a}_{\mathbf{k}}+\sqrt{\rho
_{0}}\sum_{\mathbf{k}\neq 0}\left( \hat{a}_{\mathbf{k}}^{\dagger }V_{-%
\mathbf{k}}+\hat{a}_{\mathbf{k}}V_{\mathbf{k}}\right)   \notag \\
&+&\frac{1}{2}g(t)\rho _{0}\sum_{\mathbf{k}\neq 0}\left( \hat{a}_{\mathbf{k}%
}^{\dagger }\hat{a}_{-\mathbf{k}}^{\dagger }+\hat{a}_{\mathbf{k}}\hat{a}_{-%
\mathbf{k}}\right) ,  \label{EH}
\end{eqnarray}%
where $\hat{a}_{\mathbf{k}}(\hat{a}_{\mathbf{k}}^{\dag })$ annihilates
(creates) a bosonic atom with momentum $\mathbf{k}$, and $\rho _{0}=N_{0}/V$
is the condensate density with $N_{0}$ being the number of condensed atoms.
Hamiltonian (\ref{EH}) describes the process when a pair of bosonic atoms
with momenta $\{\mathbf{k},-\mathbf{k}\}$ are annihilated through the two
body interaction (and vice versa), as well as the process when a single
particle with momenta $\mathbf{k}$ is scattered by the disordered potential
into the condensate (and vice versa).

As the system is initially prepared at the ground state of Hamiltonian $H_{
\text{eff}}$ (\ref{EH}) with $g=g_{i}$, the quench from $g_{i}$ to $g_{f}$ will bring
the system to the nonequilibrium. For the quadratic Hamiltonian (\ref{EH}),
the nonequilibrium dynamics can be exactly described as $|\Phi (t)\rangle
=\Pi _{\mathbf{k}}U_{\mathbf{k}}(t)|\Phi (0)\rangle $ with $|\Phi (0)\rangle
$ and $|\Phi (t)\rangle $ being the many-body wavefunction before and after
the quench, respectively, and $U_{\mathbf{k}}\left( t\right) $ represents
the evolution operator for each momenta $\mathbf{k}$. By noticing that
Hamiltonian $H_{\text{eff}}$ with $g=g_{f}$ contains the operators of $K_{0}(%
\mathbf{k})=(\hat{a}_{\mathbf{k}}^{\dagger }\hat{a}_{\mathbf{k}}+\hat{a}_{-%
\mathbf{k}}^{\dagger }\hat{a}_{-\mathbf{k}})/2$, $K_{+}(\mathbf{k})=\hat{a}_{%
\mathbf{k}}^{\dagger }\hat{a}_{-\mathbf{k}}^{\dagger }$ and $K_{-}(\mathbf{k}%
)=\hat{a}_{\mathbf{k}}\hat{a}_{-\mathbf{k}}$ which form the generators of
SU(1,1) Lie algebra \cite{Quench1,Truax}, we can obtain $U_{\mathbf{k}}(t)$
as
\begin{eqnarray}
U_{\mathbf{k}}\left( t\right)  &=&e^{\alpha _{\mathbf{k}}^{\ast }\left(
t\right) \hat{a}_{\mathbf{k}}^{\dagger }}e^{\alpha _{\mathbf{k}}\left(
t\right) \hat{a}_{\mathbf{k}}}\exp \left[ \beta _{0}\left( {\mathbf{k}}%
,t\right) K_{0}\left( \mathbf{k}\right) +i\phi _{\mathbf{k}}\left( t\right) %
\right]   \notag \\
&\times &\exp \left[ \beta _{+}\left( \mathbf{k},t\right) K_{+}\left(
\mathbf{k}\right) \right] \exp \left[ \beta _{-}\left( \mathbf{k},t\right)
K_{-}\left( \mathbf{k}\right) \right] .  \label{Uk}
\end{eqnarray}%
Here, $\phi _{\mathbf{k}}(t)$ is a trivial phase and
\begin{eqnarray}
\alpha _{\mathbf{k}}(t) &=&-\frac{\sqrt{\rho _{0}}}{\omega _{\mathbf{k}}}V_{%
\mathbf{k}}\left( \left\vert u_{\mathbf{k}}\right\vert ^{2}+2u_{\mathbf{k}%
}v_{\mathbf{k}}^{\ast }+\left\vert v_{\mathbf{k}}\right\vert ^{2}\right) ,
\notag \\
\beta _{+}(\mathbf{k},t) &=&v_{\mathbf{k}}^{\ast }(t)/u_{\mathbf{k}}^{\ast
}(t),\beta _{-}(\mathbf{k},t)=-v_{\mathbf{k}}(t)/u_{\mathbf{k}}^{\ast }(t),
\notag \\
\beta _{0}(\mathbf{k},t) &=&-2\ln {u_{\mathbf{k}}^{\ast }(t)},
\end{eqnarray}%
are expressed in terms of the disorder potential $V_{\mathbf{k}}$ and the
time-dependent Bogoliubov amplitudes $u_{\mathbf{k}}(t)$ and $v_{\mathbf{k}%
}(t)$ which are determined from
\begin{eqnarray}
\left(
\begin{array}{c}
u_{\mathbf{k}}(t) \\
v_{\mathbf{k}}(t)%
\end{array}%
\right)  &=&\Bigg[\cos (\omega _{\mathbf{k}}^{f}t)\hat{I}-i\frac{\sin
(\omega _{\mathbf{k}}^{f}t)}{\omega _{\mathbf{k}}^{f}}\times   \notag \\
&&\left(
\begin{array}{cc}
\epsilon _{\mathbf{k}}+g_{f}\rho _{0} & g_{f}\rho _{0} \\
-g_{f}\rho _{0} & -(\epsilon _{\mathbf{k}}+g_{f}n\rho _{0})%
\end{array}%
\right) \Bigg]\left(
\begin{array}{c}
u_{\mathbf{k}}\left( 0\right)  \\
v_{\mathbf{k}}\left( 0\right)
\end{array}%
\right) ,  \label{UV}
\end{eqnarray}%
with $u_{\mathbf{k}}\left( 0\right) (v_{\mathbf{k}}\left( 0\right) )=\pm
\sqrt{\left[ (\epsilon _{\mathbf{k}}+\rho _{0}g_{i})/\omega _{\mathbf{k}%
}^{i}\pm 1\right] /2}$ and $\omega _{\mathbf{k}}^{f(i)}=\sqrt{\epsilon _{%
\mathbf{k}}\left( \epsilon _{\mathbf{k}}+2\rho _{0}g_{f(i)}\right) }$ and $%
\epsilon _{\mathbf{k}}=\hbar ^{2}k^{2}/2m$. Note that $\left\vert u_{\mathbf{%
k}}\left( t\right) \right\vert ^{2}-\left\vert v_{\mathbf{k}}\left( t\right)
\right\vert ^{2}=1$ is always satisfied during the time evolution. For
self-consistency, hereafter we limit ourselves in the regime where the time
dependence of $\rho _{0}$ can be ignored \cite{Natu2013}.

\textit{Quantum depletion after quench.--} We are now well equipped to study the
time evolution of the non-condensed fraction $\rho_{ex}(t)=\langle
\Phi(0)|\sum_{\mathbf{k}} \hat{a}_{\mathbf{k}}^\dag \hat{a}_{\mathbf{k}%
}|\Phi(0)\rangle$ of the considered system, given that the initial condition that
$|\Phi(0)\rangle$ is the ground state of $H_i$. Straightforward derivation
using Eq. (\ref{Uk}) yields \cite{UedaBook}
\begin{equation}  \label{rho}
\rho_{ex}(t)=\sum_{\mathbf{k}}|\upsilon_{\mathbf{k}}(t)|^2.
\end{equation}
By substituting Eq. (\ref{UV}) into Eq. (\ref{rho}), we can arrive at
\begin{widetext}
\begin{eqnarray}
\frac{\rho_{ex} (t)}{(3\pi^2 \zeta^3)^{-1}}& = & 1+\frac{3\pi\tilde{R}_0}{2+2\sqrt{\tilde{ g}}}-3\sqrt{2}\int_{0}^{\infty}d{k}{\frac{\tilde{g}\left(1-\tilde{g}\right){k}^{2}\sin^{2}\left(\sqrt{{k}^{2}\left({k}^{2}+2\tilde{g}\right)}{t}\right)}{\left({k}^{2}+2\tilde{g}\right)\sqrt{{k}^{2}\left({k}^{2}+2\right)}}}\nonumber\\
&+&6\sqrt{2}\tilde{R}_0\left(1-\tilde{g}\right)\int_{0}^{\infty}d{k}{\frac{2\left({k}^{4}+k^2+\tilde{g}k^2\right)\sin^{2}(\sqrt{{k}^{2}\left({k}^{2}+2\tilde{g}\right)}{t})+\tilde{g}(1-\tilde{g})\sin^{2}(2\sqrt{{k}^{2}\left({k}^{2}+2\tilde{g}\right)}{t})}{\left({k}^{2}+2\right)\left({k}^{2}+2\tilde{g}\right)^{3}}},\label{OneBody}
\end{eqnarray}
\end{widetext}
where $\zeta=\hbar/\sqrt{mg_i\rho_0}$ is the initial healing length, and we
have introduced the dimensionless parameter
\begin{equation}
\tilde{R}_0=\rho_0R_0/(g_i\rho_0)^2  \label{R0tilde}
\end{equation}
to characterise the relative disorder strength. Note that, for vanishing
disorder $\tilde{R}_0=0$, Equation (\ref{OneBody}) agrees exactly with the
corresponding result in Ref. \cite{Natu2013}; whereas for $\tilde{g}=1$ (no
quench), the system simply remains in the ground state $|\Phi(0)\rangle$ with
the depletion $\rho_{ex}= (3\pi^2\zeta^3)\left[1+3\pi\tilde{R}_0/4\right]$
as in Ref. \cite{Huang1992}. Therefore, the last two terms in Eq. (\ref%
{OneBody}) presents the combined effect of the interaction quench $\tilde{g}%
\neq 1$ and disorder $\tilde{R}_0$ on the condensate depletion in the
non-equilibrium regime.

We are interested in the asymptotic behavior of $\rho _{ex}(t)$ at the long
time after the quench. To comprehensively reveal the roles of $\tilde{R}_{0}$
and $\tilde{g}$, we consider three cases for numerical analysis of
Eq. (\ref{OneBody}), as illustrated in Fig. \ref{Condensatedensity}. (i)
Firstly, we show how the quench strength $\tilde{g}$ affects the asymptotic
depletion. As such, we fix $\tilde{R}_{0}=0$ and calculate $\rho _{ex}(t)$
(blue solid line), which is compared to the corresponding equilibrium
depletion for a BEC with $g_{f}$ (blue dashed line). (ii) Secondly, to extract
the role of disorder, we fix $\tilde{g}=0$ that corresponds to a quench to a
non-interacting BEC (red solid line) with disorder, and then compare $\rho
_{ex}(t)$ with the corresponding equilibrium value (red dashed line). (iii)
Finally, the combined effects of disorder and quenched interaction is
illustrated by the black solid curve. In all cases, we have found
enhanced depletion in the asymptotic steady state compared to the
corresponding case at zero temperature. This indicate that the ability of
disorder or interaction to deplete the condensate is magnified in the
non-equilibrium scenario.

Compared to the equilibrium disordered 3D BEC, the increased depletion in
the steady state of the corresponding system under an interaction quench can
be qualitatively understood in terms of the Loschmidt echo $L(t)=|\langle
\Psi _{0}(t)|\Psi (t)\rangle |^{2}$ \cite{LE1,LE2}. The Loschmidt echo has
been intensively studied recently in quenched systems. The connection
between the condensate depletion and the Loschmidt echo is best illustrated
in the case of $\tilde{R}=0$ in the quadratic Hamiltonian $H_{\text{eff}}$,
when the Loschmidt echo can be calculated as $L(t)=1/\Pi _{\mathbf{k}}|u_{%
\mathbf{k}}|^{2}$ \cite{Quench1}. By using Eq. (\ref{rho}) and $|u_{\mathbf{k%
}}|^{2}=1+|\upsilon _{\mathbf{k}}|^{2}$, we estimate $L(t)\approx 1/(1+\rho
_{ex}(t))$ (after ignoring higher order terms in $|u_{\mathbf{k}}|^{2}$ and $%
|\upsilon _{\mathbf{k}}|^{2}$ ). Then, building on the square relation $%
L_{sq}(t\rightarrow \infty )=L_{ad}^{2}(t\rightarrow \infty )$ established
in Ref. \cite{Quench1,SQLE,Heyl2013}, which connects the steady-state
Loschmidt echo in a sudden quench ($L_{sq}$) to that of an adiabatic
interaction change ($L_{ad}$), we can estimate $\rho _{ex}^{sq}(t\rightarrow
\infty )=(1+\rho _{ex}^{ad})^{2}-1$. To see how this formula fits, we input
the adiabatic value $\rho _{ex}^{eq}(t)/\rho _{ex,eq}=0.196$ (obtained from
the dashed blue line in Fig. \ref{Condensatedensity} a) and calculate the
sudden quench depletion as $\rho _{ex}^{sq}(t\rightarrow \infty )/\rho
_{ex,eq}=0.43$, which agrees fairly well with the numerical results in the
steady state (solid blue line in Fig. \ref{Condensatedensity} a).

\begin{figure}[thb]
\includegraphics[width=8.0cm]{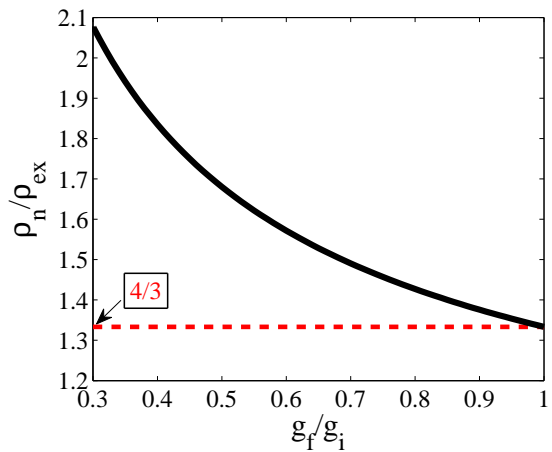}
\caption{(Color online) Quench enhanced ratio between  normal density $\rho_n$ in Eq. (\ref{rhon}) and quantum depletion $\rho_{ex}(t\rightarrow \infty)$ in Eq. (\ref{OneBody}) via $g_f/g_i$.
The dashed line corresponds to the value of $\rho_n/\rho_{ex}=4/3$ first obtained in Ref. \cite{Huang1992}.}
\label{Boseglass}
\end{figure}

\textit{Superfluid depletion after quench.--} The superfluid component and
the normal fluid component can be clearly distinguished from their response
to a slow rotation: normal fluid rotates but superfluid does not. This is
the essential concept behind typical experimental schemes to measure the
superfluid density in an atomic Bose gas \cite%
{Cooper2010,Ho2009,Carusotto2011}. While the techniques to generate rotations
in a Bose gas differ in various schemes, the key quantity measured boils
down to the current-current response function $\chi _{i,j}\left( \mathbf{r}t,%
\mathbf{r}^{\prime }t^{\prime }\right) =\langle \left[ J_{i}(\mathbf{r}%
,t),J_{j}(\mathbf{r}^{\prime },t^{\prime })\right] \rangle $ ($i,j=x,y,z$)
with $J_{i}\left( \mathbf{r},t\right) =\hbar /(2i)[\hat{\Psi}^{\dagger }(%
\mathbf{r})\nabla \hat{\Psi}(\mathbf{r})-\hat{\Psi}(\mathbf{r})\nabla \hat{%
\Psi}^{\dagger }(\mathbf{r})]$ being the current density of system and $%
\langle ...\rangle $ averaged with the initial ground state $|\Psi
(0)\rangle $. Particularly, the superfluid density $\rho _{s}$ corresponds
to the response to the irrotational (longitudinal) part of the perturbation;
whereas, the normal fluid density $\rho _{n}$ describes the transverse
response, in an isotropic translationally invariant system, we have
\begin{equation}
m\chi _{ij}(\mathbf{q}\rightarrow 0,\omega )=\rho _{s}\frac{q_{i}q_{j}}{q^{2}%
}+\rho _{n}\delta _{ij}.
\end{equation}%
Based on the current experimental approaches to measure the superfluid
response, we have calculated $\rho _{n}=\chi _{zz}(\mathbf{q}\rightarrow
0,\omega )$ with $J_{\mathbf{q}}^{z}=\hbar /(2m)\sum_{\mathbf{k}%
}(k_{z}+q_{z}/2)\hat{a}_{\mathbf{k}}^{\dagger }\hat{a}_{\mathbf{k+q}}$ for
the steady state of the considered BEC after the interaction quench (as the
system is isotropic, we are free to choose the slow rotation around the $z$
axis ). Then, by using Eqs. (\ref{Uk}) and (\ref{UV}) \cite{UedaBook}, we
derive the normal fluid density in the steady state of the quenched system
as
\begin{eqnarray}
\frac{\rho _{n}}{(3\pi ^{2}\zeta ^{3})^{-1}} &=&\frac{2\pi \tilde{R}_{0}}{9%
\sqrt{\tilde{g}^{3}}}\Bigg[\left( 7\tilde{g}-2\right) \frac{\mathtt{_{2}F_{1}%
}(\frac{1}{2},\frac{1}{2};1;\frac{(\tilde{g}-1)^{2}}{\tilde{g}^{2}})}{(%
\tilde{g}-1)/\tilde{g}}  \notag \\
&-&5\frac{\mathtt{_{2}F_{1}}(-\frac{1}{2},\frac{1}{2};1;\frac{(\tilde{g}%
-1)^{2}}{\tilde{g}^{2}})}{(\tilde{g}-1)/\tilde{g}}\Bigg],  \label{rhon}
\end{eqnarray}
with $\mathtt{_{2}F_{1}(a,b;c;x)}$ being the hypergeometric function \cite%
{Mathematics}.

Now, with both the condensate depletion and the normal fluid density at hand
for the steady state of the considered system, we plot $\rho_{n}/\rho_{ex}$
as a function of the quench strength $\tilde{g}=g_f/g_i$, as illustrated in
Fig. \ref{Boseglass}. To set a reference point, we have shown that in the
limit $\tilde{g}\rightarrow 1$ (red dashed line), Eq. (\ref{rhon}) exactly
recovers the celebrated ratio $\rho_n/\rho_{ex}=4/3$ in Ref. \cite{Huang1992}
for a ground-state BEC with weak disorder. In comparison, the quench effect
(black solid line) gives rise to significantly enhanced ratio $\rho_n/\rho_{ex}$, the stronger the quench is, the higher value of 
$\rho_n/\rho_{ex}$ is found, which can even approach $2$. Moreover, we have
found that the ratio does not depend on the individual absolute value of $g_f$ and $g_i$, but rather on their relative strength $\tilde{g}=g_f/g_i$.
Figure \ref{Boseglass} shows that the ability of disorder to deplete more
superfluid than the condensate is remarkably amplified when combined with
the quench effect, which presents the major result of this work. In
principle, the value of $\rho_s/\rho_0$ can be further suppressed by
repeating the process of sudden quench in the interaction: quench from $g_i$
to $g_f$, holding time $t$ and then adiabatically change interaction from $%
g_f$ to $g_i$, then quench interaction $n$ times (bang-bang protocol). It's
highly expected that the disordered BEC can be quenched into the regime of $%
\rho_s/\rho_0\ll 1$.

\begin{figure}[t]
\includegraphics[width=8.0cm]{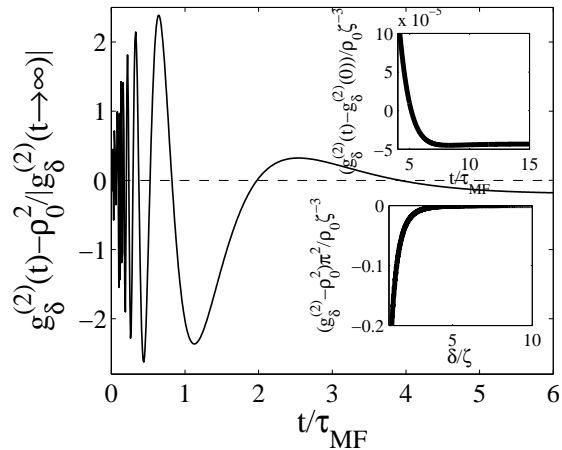}
\caption{(Color online) Quench dynamics of density-density correlation
function, which develops in an oscillatory manner and rapidly saturate at
times $t-\protect\delta/c>\protect\tau_{MF}$ with $\protect\delta$ being the
separation between points, $\protect\delta=|\mathbf{r}-\mathbf{r}^{^{\prime
}}|$. The top inset: long-time density correlation function via time. The
bottom inset: long-time behavior of density-density correlation function via the
separation between points. The parameters are given as $\tilde{\delta}=\delta/\zeta=4$, $\tilde{g}=0.6$, and $\tilde{R}_0=0.1$.}
\label{FigureTwoBody}
\end{figure}

\textit{Discussion and Conclusion.---} We have shown that, by quenching in
the interaction of a 3D BEC in disordered potential, the system can relax to
a steady state where the ratio $\rho _{n}/\rho _{ex}>4/3$ can be achieved.
This indicates the quench effect can significantly enhance the ability of
disorder to deplete the superfluid more than the condensate, and therefore,
suggests an alternative way of engineering Bose Glass in the dynamic regime.
Central to testing our observation is the experimental ability to measure
the condensate depletion $\rho _{ex}$ and normal fluid density $\rho _{n}$.
In typical experiments, the-state-of-art higher-resolution imaging
techniques allow one to probe the time-dependent quantum depletion $\rho
_{ex}$ \cite{Xu,Simon2011,Weitenberg2011}, while the experimental schemes
reported in Refs. \cite{Cooper2010,Ho2009,Carusotto2011} can be used to
measure $\rho _{n}$ in both equilibrium and nonequilibrium regimes. One
concern that may arise is related to how fast the considered system can
relax to the steady state. Figure \ref{Condensatedensity} indicates the
rapid relaxation of the one-body correlation. Let us further analyze the
quench dynamics of the two-body correlation function, a quantity directly
relevant for the the Bragg spectroscopy \cite{BraggOL1,BraggOL2} that has
been a routine technique for studying excitations of a Bose gas. The
two-body correlation function is defined as $g^{\left( 2\right) }\left(
t\right) =\sum_{\mathbf{q}}e^{i\mathbf{q}\cdot \left( \mathbf{r}-\mathbf{r}%
^{\prime }\right) }\left\langle \rho _{\mathbf{q}}\left( t\right) \rho _{-%
\mathbf{q}}\left( t\right) \right\rangle $ with $\rho _{\mathbf{q}}\left(
t\right) =\sum_{\mathbf{k}}\hat{a}_{\mathbf{k}+\mathbf{q}}^{\dagger }\left(
t\right) \hat{a}_{\mathbf{k}}\left( t\right) $ being the density operator.
Following similar procedures \cite{{Natu2013}}, we calculate the
correlation function from Eq. (\ref{Uk}) and obtain
\begin{widetext}
\begin{eqnarray}
g_{\delta}^{\left(2\right)}\left(t\right) & = & \rho_{0}^{2}+\frac{\rho_{0}}{\pi^{2}\tilde{\delta}\zeta^{3}}\int kd{k}\sin\left(\sqrt{2}{k}\tilde{\delta}\right)\Big[\frac{k}{\sqrt{k^{2}+2}}-2\left(\tilde{g}-1\right)\frac{{k}\sin^{2}\left(\sqrt{{k}^{2}\left({k}^{2}+2\tilde{g}\right)}{t}\right)}{\sqrt{{k}^{2}+2}\left({k}^{2}+2\tilde{g}\right)}-1\Big]\nonumber\\
 &+&\frac{4\rho_0}{\pi^{2}\tilde{\delta}\zeta^{3}}\tilde{R}_{0}\int d{k}\frac{{k}\sin\left(\sqrt{2}{k}\tilde{\delta}\right)}{\left({k}^{2}+2\right)\left({k}^{2}+2\tilde{g}\right)}\left\{ \cos^{2}\left(\sqrt{{k}^{2}\left({k}^{2}+2\tilde{g}\right)}{t}\right)+\frac{{k}^{2}+2}{{k}^{2}+2\tilde{g}}\sin^{2}\left(\sqrt{{k}^{2}\left({k}^{2}+2\tilde{g}\right)}{t}\right)\right\} ^{2},\label{TwoBody}
\end{eqnarray}
\end{widetext}with the dimensionless parameter $\tilde{\delta}=\delta /\zeta
$ for $\delta =|\mathbf{r}-\mathbf{r}^{\prime }|$. Again, for vanishing
disorder $\tilde{R}_{0}=0$, Equation (\ref{OneBody}) agrees with Ref. \cite%
{Natu2013}; whereas for $\tilde{g}=1$, our result is consistent with Ref.
\cite{Huang1992}. The time evolution of $g^{(2)}$ is presented in Fig. \ref%
{FigureTwoBody}, which shows a rapid relaxation to a finite value on a time
scale $t\sim 3\tau _{MF}$. Both the rapid relaxation of one-body matrix in
Fig. \ref{Condensatedensity} and the two-body correlation function in Fig. %
\ref{FigureTwoBody} suggest that, after the 3D disordered BEC is brought out
of equilibrium by a quantum quench in the interaction, it relaxes to a
steady state on a time-scale within the experimental reach.

We thank Biao Wu and Li You for stimulating discussions. This work is
supported by the NSF of China (Grant Nos. 11004200 and 11274315). Y. H.
also acknowledges support by the Austrian Science Fund (FWF) through SFB F40
FOQUS.

\end{document}